\documentclass[conference]{IEEEtran}
\IEEEoverridecommandlockouts
\usepackage{cite}
\usepackage{amsmath,amssymb,amsfonts}
\usepackage{algorithmic}
\usepackage{graphicx}
\usepackage{textcomp}
\usepackage{xcolor}
\usepackage[nolist]{acronym}
\usepackage{balance}
\usepackage{longtable}
\usepackage{relsize}
\usepackage{threeparttable}
\usepackage{cite}
\usepackage[normalem]{ulem}
\usepackage{comment,subfigure}
\usepackage{color,soul}
\usepackage{multicol,multirow,xcolor,colortbl}
\usepackage{makecell}
\def\BibTeX{{\rm B\kern-.05em{\sc i\kern-.025em b}\kern-.08em
    T\kern-.1667em\lower.7ex\hbox{E}\kern-.125emX}}

\newcommand{\sinr}{\rho}
\newcommand{\Prx}{P_\text{r}}
\newcommand{\Pn}{P_\text{n}}
\newcommand{\PI}{P_\text{I}}
\newcommand{\setI}{\mathcal{I}}
\newcommand{\Tcbr}{T_\text{CBR}}

\newcommand{\Tbusynet}{T_\text{net}}
\newcommand{\netcbr}{\gamma}
\newcommand{\Pcbr}{\theta_\text{CBR}}
\newcommand{\Ntx}{N_\text{rep}}
\newcommand{\Ntxav}{\overline{N}_\text{rep}}

\newcommand{\thrcbr}[1]{{\gamma^*_{#1}}}
\newcommand{\metric}[2]{{\psi}_{#2}({#1})}
\newcommand{\averageN}{n^{\#}}
\newcommand{\vInBox}[3]{{#1}\left[\right._{#2}^{#3}}

\newcommand{\red}[1] {\textcolor[rgb]{1.0,0.0,0.0}{{#1}}}

\begin{acronym} 
\acro{3GPP}{Third Generation Partnership Project}
\acro{2D}{two-dimensional}
\acro{2G}{second generation}
\acro{3G}{third generation}
\acro{4G}{fourth generation}
\acro{5G}{fifth generation}
\acro{6G}{sixth generation}
\acro{5GAA}{5G Automotive Association}
\acro{5GS}{5G system}
\acro{AF}{application function}
\acro{AIFS}{arbitration inter-frame space}
\acro{AMC}{adaptive modulation and coding}
\acro{AoI}{age of information}
\acro{AS}{application server}
\acro{AWGN}{additive white Gaussian noise}
\acro{BEP}{beacon error probability}
\acro{BM-SC}{Broadcast Multicast Service Center}
\acro{BP}{beacon periodicity}
\acro{B-CSA}{broadcast \ac{CSA}}
\acro{BSM}{basic safety message}
\acro{BSS}{basic service set}
\acro{BS}{base station}
\acro{C-ITS}{cooperative-intelligent transport systems}
\acro{C-NOMA}{Cooperative \ac{NOMA}}
\acro{C-V2X}{cellular-\ac{V2X}}
\acro{CA}{collision avoidance}
\acro{CAM}{cooperative awareness message}
\acro{CAV}{connected and autonomous vehicle}
\acro{CBR}{channel busy ratio}
\acro{CCDF}{complementary cumulative distribution function}
\acro{CDF}{cumulative distribution function}
\acro{CDMA}{code-division multiple access}
\acro{CoMP}{coordinated multi-point}
\acro{CN}{core network}
\acro{CP}{cyclic prefix}
\acro{CP-OFDM}{cyclic prefix orthogonal frequency-division multiplexing}
\acro{CPM}{collective perception message}
\acro{CR}{channel occupancy ratio}
\acro{CRDSA}{contention resolution diversity slotted ALOHA}
\acro{CSA}{coded-slotted ALOHA}
\acro{CSD}{cyclic shift diversity}
\acro{CSI}{channel state information}
\acro{CSIT}{channel state information at the transmitter}
\acro{CSIR}{channel state information at the receiver}
\acro{CSMA/CA}{carrier sense multiple access with collision avoidance}
\acro{CSSR}{candidate single-subframe resource}
\acro{D2D}{device-to-device}
\acro{DCC}{decentralized congestion control}
\acro{DCF}{distributed coordination function}
\acro{DCI}{downlink control
information}
\acro{DCM}{dual carrier modulation}
\acro{DENM}{decentralized environmental notification message}
\acro{DMRS}{demodulation reference signal}
\acro{DRX}{discontinuous reception}
\acro{DSRC}{dedicated short range communication}
\acro{DOT}{Department of Transportation}
\acro{EC}{European Commission}
\acro{ECP}{extended cyclic prefix}
\acro{EDCA}{enhanced distributed coordination access}
\acro{EDGE}{Enhanced Data rates for GSM Evolution}
\acro{eMBMS}{evolved Multicast Broadcast Multimedia Service}
\acro{eMMB}{enhanced mobile broadband}
\acro{eNodeB}{evolved NodeB}
\acro{EPC}{Evolved Packet Core}
\acro{EPS}{Evolved Packet System}
\acro{ETSI}{European Telecommunications Standards Institute}
\acro{eV2X}{enhanced V2X}
\acro{FCC}{Federal Communications Commission}
\acro{FCD}{floating car data}
\acro{FD}{in-band full-duplex}
\acro{FDD}{frequency division duplex}
\acro{FDMA}{frequency division multiple access}
\acro{FEC}{forward error correction}
\acro{FR1}{frequency range 1}
\acro{FR2}{frequency range 2}
\acro{GNSS}{global navigation satellite system}
\acro{GPRS}{general packet radio service }
\acro{GPS}{global positioning system}
\acro{HARQ}{hybrid automatic repeat request}
\acro{HD}{half-duplex}
\acro{HSDPA}{High Speed Downlink Packet Access}
\acro{HSPA}{High Speed Packet Access}
\acro{IBE}{in-band emission}
\acro{ICI}{Inter-carrier interference}
\acro{IDMA}{interleave-division multiple access}
\acro{IM}{index modulation}
\acro{IP}{Internet Protocol}
\acro{IPG}{inter-packet gap}
\acro{IR}{incremental redundancy}
\acro{ISI}{inter-symbol interference}
\acro{ITS}{intelligent transport system}
\acro{i.i.d.}{independent identically distributed}
\acro{KPI}{key performance indicator}  
\acro{LDPC}{low-density parity-check}
\acro{LDS}{low-density spreading}
\acro{LIDAR}{light detection and ranging}
\acro{LOS}{line-of-sight}
\acro{LTE}{long term evolution}  
\acro{LTE-D2D}{\ac{LTE} with \ac{D2D} communications}  
\acro{LTE-V2X}{long-term-evolution-vehicle-to-anything} \acro{LTE-V2V}{\ac{LTE}-\ac{V2V}}
\acro{MAC}{medium access control}
\acro{MBMS}{Multicast Broadcast Multimedia Service}
\acro{MBMS-GW}{MBMS Gateway}
\acro{MBSFN}{MBMS Single Frequency Network}
\acro{MCE}{Multi-cell Coordination Entity}
\acro{MEC}{mobile edge computing}
\acro{MCM}{maneuver coordination message}
\acro{MCS}{modulation and coding scheme}
\acro{MIMO}{multiple input multiple output}
\acro{ML}{machine learning}
\acro{MRC}{maximum ratio combining}
\acro{MRD}{maximum reuse distance}
\acro{MUD}{multi-user detection}
\acro{NEF}{network exposure function}
\acro{NGV}{Next Generation V2X}
\acro{NHTSA}{National Highway Traffic Safety Administration}
\acro{NLOS}{non-line-of-sight}
\acro{NOMA}{non-orthogonal multiple access}
\acro{NOMA-MCD}{\ac{NOMA}-mixed centralized/distributed}
\acro{NR}{new radio}
\acro{OBU}{on-board unit}
\acro{OCB}{outside of the context of a basic service set}
\acro{OEM}{Original equipment manufacturer}
\acro{OFDM}{orthogonal frequency-division multiplexing}
\acro{OFDMA}{orthogonal frequency-division multiple access}
\acro{OMA}{orthogonal multiple access}
\acro{OTFS}{orthogonal time frequency space}
\acro{pdf}{probability density function}
\acro{PAPR}{peak to average power ratio}
\acro{PCF}{policy control function}
\acro{PCM}{platoon control message}
\acro{PD}{packet delay}
\acro{PEP}{pairwise error probability} 
\acro{PER}{packet error rate}
\acro{PIAT}{packet inter-arrival time}
\acro{PRB}{physical resource block}
\acro{PSCCH}{Physical Sidelink Control Channel}
\acro{PSFCH}{Physical Sidelink Feedback channel}
\acro{PSSCH}{Physical Sidelink Shared channel}
\acro{PHY}{physical}
\acro{PL}{path loss}
\acro{PLMN}{Public Land Mobile Network}
\acro{PPPP}{ProSe Per-Packet Priority}
\acro{PPPR}{ProSe Per-Packet Reliability}
\acro{ProSe}{Proximity-based Services}
\acro{PRR}{packet reception ratio}
\acro{QAM}{quadrature amplitude modulation}
\acro{QC-LDPC}{quasi-cyclic \ac{LDPC}}
\acro{QoS}{quality of service}
\acro{QPSK}{quadrature phase shift keying}
\acro{RAN}{Radio Access Network}
\acro{RAT}{radio access technology}
\acro{RF}{radio frequency}
\acro{RIS}{reconfigurable intelligent  surface}
\acro{RL}{reinforcement learning}
\acro{RR}{radio resource}
\acro{RRC}{radio resource control}
\acro{RRI}{resource reservation interval}
\acro{RSRP}{reference signal received power}
\acro{RSSI}{Received Signal Strength Indicator}
\acro{RSU}{road side unit} 
\acro{SAE}{Society of Automotive Engineers}
\acro{SAI}{Service Area Identifier}
\acro{SB-SPS}{sensing-based semi-persistent scheduling}
\acro{SC-FDMA}{single carrier frequency division multiple access}
\acro{SC-PTM}{Single Cell Point To Multipoint}
\acro{SCS}{subcarrier spacing}
\acro{SCMA}{sparse-code multiple access}
\acro{SI}{self-interference}
\acro{SIC}{successive interference cancellation}
\acro{SIFS}{short inter-frame space}
\acro{S-UE}{scheduling UE}
\acro{SCI}{sidelink control information}
\acro{SDN}{Software-defined networking}
\acro{SFFT}{symplectic finite Fourier transform}
\acro{SHINE}{simulation platform for heterogeneous interworking networks}
\acro{SINR}{signal-to-interference-plus-noise ratio}
\acro{SNR}{signal-to-noise ratio}
\acro{SRS}{sounding reference signal}
\acro{STBC}{space time block codes}
\acro{SV}{smart vehicle}
\acro{TB}{transport block}
\acro{TBC}{time before change}
\acro{TBE}{time before evaluation}
\acro{TDD}{time-division duplex}
\acro{TDMA}{time-division multiple access}
\acro{TR}{technical report}
\acro{TS}{technical specification}
\acro{TTI}{transmission time interval}
\acro{UAV}{unmanned aerial vehicle}
\acro{UD}{Update delay}
\acro{PSDU}{Physical Layer Convergence 
Protocol Service 
Data Unit}
\acro{UMTS}{universal mobile telecommunications system}
\acro{URLLC}{ultra-reliable and ultra-low latency communications}
\acro{USIM}{Universal Subscriber Identity Module}
\acro{UTDOA}{uplink time difference of arrival}
\acro{V2C}{vehicle-to-cellular} 
\acro{V2N}{vehicle-to-network}
\acro{V2I}{vehicle-to-infrastructure} 
\acro{V2P}{vehicle-to-pedestrian}
\acro{V2R}{vehicle-to-roadside}
\acro{V2V}{vehicle-to-vehicle} 
\acro{V2X}{Vehicle-to-everything} 
\acro{VAM}{VRU awareness message}
\acro{VRU}{vulnerable road user}
\acro{VUE}{vehicular user equipment}
\acro{WAVE}{wireless access in vehicular environment}
\acro{WBSP}{wireless blind spot probability}
\acro{PPDU}{physical layer protocol data unit}
\acro{LLR}{log-likelihood ratio}
\end{acronym}

\begin{document}

\title{Adaptive Repetitions Strategies\\in IEEE 802.11bd 
\thanks{We would like to thank the China Scholarship Council that is supporting Wu Zhuofei during his visiting scholarship at the University of Bologna.}
}


\author{\IEEEauthorblockN{Wu Zhuofei\IEEEauthorrefmark{1}\IEEEauthorrefmark{2}, Stefania Bartoletti \IEEEauthorrefmark{3}, Vincent Martinez\IEEEauthorrefmark{4}, and Alessandro Bazzi\IEEEauthorrefmark{2}}

\IEEEauthorblockA{\IEEEauthorrefmark{1}\textit{Harbin Engineering University}, Harbin, China, wzfhrb@hrbeu.edu.cn\\}
\IEEEauthorblockA{\IEEEauthorrefmark{2}\textit{WiLab, CNIT / DEI, University of Bologna}, Bologna, Italy, \{wu.zhuofei, alessandro.bazzi\}@unibo.it\\}
\IEEEauthorblockA{\IEEEauthorrefmark{3}\textit{IEIIT/CNR, CNIT}, Bologna, Italy,
stefania.bartoletti@ieiit.cnr.it}
\IEEEauthorblockA{\IEEEauthorrefmark{4}\textit{NXP}, Toulouse, France, vincent.martinez@nxp.com 
}

}

\markboth{Journal of \LaTeX\ Class Files,~Vol.~14, No.~8, August~2021}%
{Shell \MakeLowercase{\textit{et al.}}: A Sample Article Using IEEEtran.cls for IEEE Journals}

\maketitle

\begin{abstract}
A new backward compatible WiFi amendment is under development by the IEEE bd Task Group towards the so-called IEEE 802.11bd, which includes the possibility to transmit up to three repetitions of the same packet. This feature increases time diversity and enables the use of \ac{MRC} at the receiver to improve the probability of correct decoding. In this work, we first investigate the packet repetition feature and analyze how it looses its efficacy increasing the traffic as an higher number of transmissions may augment the channel load and collision probability. Then, we propose two strategies for adaptively selecting the number of transmissions leveraging on an adapted version of the \ac{CBR}, which is measured at the transmitter and is an indicator of the channel load. The proposed strategies are validated through network-level simulations that account for both the acquisition and decoding processes. Results show that the proposed strategies 
ensure that devices use optimal settings under variable traffic conditions. 
\end{abstract}


\acresetall

\section{Introduction}

\ac{V2X} communications will play a key role to address the so-called vision zero for traffic safety, 
where there will be no more deaths on the road. In current implementations, V2X is used by  vehicles to inform the neighbors about their own status and movements; in the future, the exchanged packets will include information about the sensed environment and coordinating manoeuvres.
After decades of researches and experiments \cite{sepulcre2022analytical, Klapez2021experimental}, the IEEE 802.11p-based \ac{V2X} is reaching the mass market.
\footnote{At the beginning of 2021, the first cars started being sold in Europe with \acp{OBU} implementing ITS-G5, based on IEEE 802.11p, as standard equipment and over 6000~km of roads were already covered by commercially distributed \acp{RSU} with the same technology (source: C-Roads at C2C-CC Forum, November 2020).} 

Given the increased interest on V2X and the necessity to improve its efficiency and reliability, a new IEEE WiFi Task Group called ``bd" was established, which is expected to publish the new amendment by the end of 2022. The so-called IEEE 802.11bd is designed with the following objectives: i) improved performance, with higher spectral efficiency (meaning higher data-rate), increased reliability, and extended range; and ii) smooth transition between IEEE 802.11p and IEEE 802.11bd, with attention to coexistence, backward compatibility, interoperability, and fairness, allowing to switch when needed between the \textit{legacy mode} (i.e., IEEE 802.11p) and the \textit{\ac{NGV} mode} (i.e., IEEE 802.11bd). 

The main features introduced by the IEEE 802.11bd are \cite{naik2019,anwar2019physical}: new \acp{MCS} based on \ac{LDPC} or \ac{DCM}; 
a known pilot sequence located periodically in the data part of the signal, which is called midamble and allows to improve the channel estimation when large packets are transmitted; 
the possibility to use 
channel bonding; 
the possibility to use mmWave bands; 
and, finally, to blindly retransmit the same packet up to three times more as a burst.

Retransmissions are in general a commonly used method to increase the \ac{PRR}, especially in the unicast communications where acknowledgments messages can be sent back from the receiver. Differently, when transmissions are performed in broadcast like in most V2X scenarios, the presence of multiple receiving stations makes it difficult to determine if and when to perform retransmissions. Nonetheless, the 3GPP in its \ac{LTE}-V2X and 5G \ac{NR}-V2X sidelink added the possibility to perform the so-called blind retransmissions, which correspond to the transmission of the same content more than once based on a decision made \textit{a priori} by the transmitter, independently on what happens at the receiver \cite{9348560,TodBarCamMolBerBaz:21}. 

\begin{figure*}[tp]
\centering
\includegraphics[width=2\columnwidth]{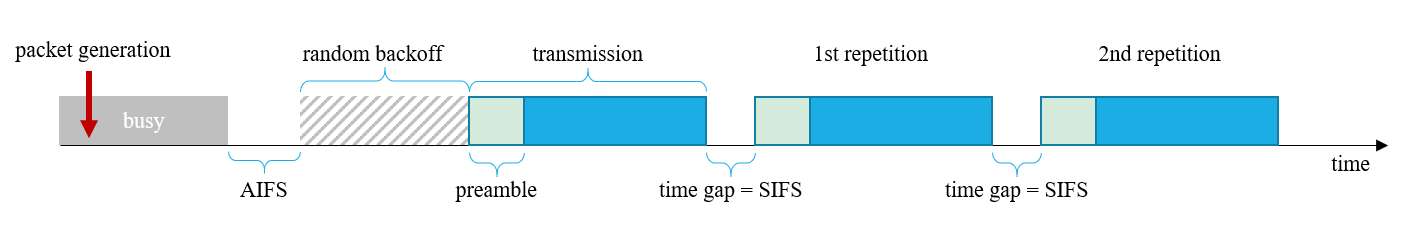}
\caption{An example of transmission with two repetitions. At the packet generation, the channel is sensed busy. Once the busy condition ends, the transmission starts after sensing the medium idle for a time interval (AIFS) followed by the random backoff. Two repetitions follow the first transmission, separated by a time gap (SIFS, which is shorter than the AIFS). Each transmission starts with a preamble indicating the presence of the packet.
} 
\label{fig:explanation}
\end{figure*}

IEEE 802.11bd introduces the possibility of blind retransmissions, 
hereafter called repetitions, which may be more effective than simply lowering the \ac{MCS}, especially in case of sporadic and strong interference or in case of high velocity conditions and deep fade. Specifically, up to three repetitions can follow the first transmission of the packet, which all carry exactly the same data  (therefore, at most four copies of the same content can be overall sent). For legacy IEEE 802.11p receivers, the probability of correct decoding of the packet is inherently improved by the increased time diversity. Additionally, IEEE 802.11bd receivers can detect the repetitions and use \ac{MRC} on the received signals to further increase the reliability of the communication.

This feature appears relevant to increase reliability and range, but comes at the cost of an increased channel load. Therefore, the choice to use it or not is subject to a trade-off which has not yet been deepened in the literature. Indeed, most of the existing studies explore the physical layer aspects of packet repetitions, i.e., looking at probability of correct reception of a generic link without interference, see e.g. \cite{anwar2019physical} and related works, whereas only  \cite{jacob2022congestion} provides early results on the impact of packet repetitions at a network level. 

In this work, we analyse the impact of the repetitions on the performance from a network point of view, considering different channel models, varying the vehicles' densities, and including the effects of preamble detection (i.e., considering that the received signals can only be combined when their preambles are detected). Then, we propose two adaptive strategies, called \textit{deterministic} and \textit{probabilistic}, to opportunistically set the number of repetitions in order to maximise the network performance. 
The proposed approach is distributed, as it leverages on an adapted version of the \ac{CBR}, which is a metric already measured at each single node. Results, obtained by the use of an open source simulator, show that the proposed approach is effective  and enables fair access to the channel.



\section{Repetitions in IEEE 802.11bd}
\label{sec:rep}



As a feature of IEEE 802.11bd, up to three repetitions can be optionally performed after the first transmission of the packet. More specifically, as exemplified in Fig.~\ref{fig:explanation}, a station accesses the channel through the \ac{CSMA/CA} mechanism which requires that the medium is idle for at least an \ac{AIFS}; the repetitions that may follow the first transmission are then separated by a time gap lasting a \ac{SIFS}, which is shorter than the AIFS, thus ensuring that the use of the channel is not released. 


In 802.11bd, each packet consists of preamble and data field. The receiver can decode the data field only if it firstly detects the preamble. At the receiver side, once the packet is correctly decoded, the subsequent repetitions are ignored. Otherwise, the receiver can store the signals of the undecoded packets for which the preambles are detected. Such stored signals can be combined through \ac{MRC} 
to improve the probability of correct reception of the packet. Note that if the preamble is not detected, the receiver is not aware of the presence of the signal and cannot perform the storage and \ac{MRC}. 

The \ac{SINR} corresponding to the $j^\text{th}$ transmission can be modeled as
\begin{equation}
    \sinr_j = \frac{{\Prx}_j}{\Pn + \sum_{i\in{\setI}_j} {\PI}_{ji}}
\end{equation}
where ${\Prx}_j$ is the received power; $\Pn$ is the average noise power; ${\setI}_j$ is the set of nodes that are interfering with the reception under examination; and ${\PI}_{ji}$ is the average power from the $i^\text{th}$ interferer.
Then, the average \ac{SINR} of the  signals combined by the MRC receiver can be modeled as 
\begin{equation}
    \sinr = \sum_{j=1}^{M}{\alpha_j\cdot\sinr}_j
\end{equation}
where $M$ is total number of transmissions, including the first transmission and the repetitions; $\alpha_j = 1$ if the preamble is detected, otherwise $\alpha_j = 0$.



\section{Repetition Strategies}
\label{sec:strategies}
In this Section, we first introduce a general approach for the setting of the number of repetitions based on the channel load and then specify two strategies (called \textit{deterministic} and \textit{probabilistic}) that can be distributively applied by the stations.

\textbf{Net channel busy ratio.} The decision on the number of repetitions $N_\text{rep}$ to be transmitted needs to be made autonomously by each station, based on its knowledge about the channel load. To this aim, the station already collects what is called \ac{CBR}\cite{etsi302663}, which is a metric used for congestion control and is defined as the average time for which the signal received from the other stations has a  power above a given threshold $\Pcbr$, and it is updated every $\Tcbr$. However, the \ac{CBR} depends on the number of repetitions, and therefore we here define the \textit{net \ac{CBR}}, also updated every $\Tcbr$, calculated as 
\begin{equation}
    \netcbr = \Tbusynet / \Tcbr
\end{equation}
where $\Tcbr$ is the duration of the observation and $\Tbusynet$ is the sum of the intervals during $\Tcbr$ where the first detected copy of a packet is received and the power of the received signal is above the threshold $\Pcbr$.  The use of the net CBR in place of the total CBR is necessary to avoid a loop triggering, in which the decision on the number of repetitions relies on a metric which in turn depends on the decision itself. 
Note also that the identification of the first copy of a packet is easily performed by the receiving station, since the repetitions that follow are spaced by a short SIFS gap.


\textbf{Approach of the strategies.} 
We propose two strategies where each station autonomously sets the number of repetitions so that a target performance metric $\metric{\netcbr{}}{\Ntx}$ is optimized. The target performance metric $\metric{\netcbr{}}{\Ntx}$ (which can represent, e.g., the maximum distance for a given \ac{PRR}) is a function of the net CBR, and also depends on the number of repetitions $N_\text{rep}$. 
%
%
%
In both the proposed strategies, the average number of repetitions $\Ntxav(\netcbr{})$ is first evaluated as a non-increasing function of the net CBR, with the net CBR domain divided into intervals. Then, the number of repetitions $\Ntx$ is determined based on $\Ntxav(\netcbr{})$. 
Following this approach, we first define the net CBR intervals and then specify the function $\Ntxav(\netcbr{})$ for the two strategies and the value of $\Ntx$ deriving from it.

\begin{figure}
\centering
\includegraphics[width=1\columnwidth]{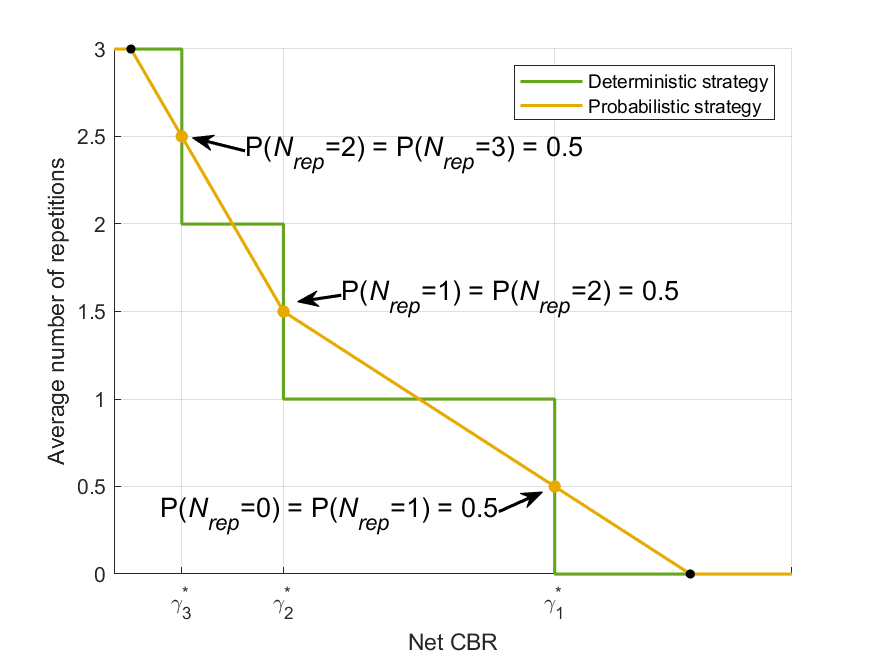}
\caption{Example of the average number of repetitions varying the net CBR with the two proposed strategies.
}
\label{fig:retrans_strategy}
\end{figure}

\textbf{Net CBR intervals.} For a fixed number of repetitions $\Ntx$, it is expected that the performance $\metric{\netcbr{}}{\Ntx}$ decreases with $\netcbr{}$. In fact, an increase of the data traffic (e.g., higher vehicle density, larger packets, more frequent packet generation), which corresponds to an increase of the net CBR $\netcbr$ usually degrades the communication performance. 
Moreover, if the channel is not congested (lower values of $\netcbr{}$), an increase of the repetitions can benefit the performance (i.e.,  $\metric{\netcbr{}}{N+1}>\metric{\netcbr{}}{N}$ for lower values of $\netcbr{}$). Differently, under a higher channel load, a further increase of repetitions can degrade the performance (i.e.,  $\metric{\netcbr{}}{N+1}<\metric{\netcbr{}}{N}$ for higher values of $\netcbr{}$). The value of $\netcbr{}$ where $\metric{\netcbr{}}{N+1}$ and $\metric{\netcbr{}}{N}$ intersect thus represents a threshold below which $N+1$ repetitions are preferable than $N$ and above which the opposite is true.

Based on such considerations, we divide the net CBR domain $[0,1]$ into $N_\text{max}+1$ intervals, 
where $N_\text{max}$ is the maximum number of repetitions. Specifically, if the CBR falls within the $i^\text{th}$ interval, the  best option is to mostly use $i-1$ repetitions. 
The $i^\text{th}$ interval is defined as $[\thrcbr{i}, \thrcbr{i-1})$, with the threshold $\thrcbr{i}$ defined as:
\begin{equation}\label{eq:thresholdi}
    \thrcbr{i} = {{\netcbr{}}^{\#}_i}\left[\right._{\thrcbr{i-1}}^{{\thrcbr{i+1}}}
\end{equation}
where we define $x\left[\right._{L}^{H} \triangleq \max{(L,\min{(x, H)})}$ 
to simplify the notation, and
where $\gamma_{0}^{*} \triangleq 1$, $\gamma_{N_\text{max}}^{*}\triangleq 0$, and 
\begin{equation}
{\netcbr{}}^{\#}_i = \min_{\netcbr{}}\{\metric{\netcbr{}}{i}>\metric{\netcbr{}}{i+1}\}\;.
\end{equation}
Equation \eqref{eq:thresholdi} guarantees that the thresholds are correctly ordered. If $\thrcbr{i}=\thrcbr{i-1}$, it simply means that transmitting $i-1$ repetitions is not convenient.




\textbf{Deterministic strategy.}
In the \textit{deterministic} strategy,
the station transmits an average number of repetitions that is a step function 
of $\netcbr{}$, i.e.
\begin{equation}
    \Ntxav(\netcbr{}) = i  \text{ where } \thrcbr{i+1}\leq \netcbr{} <\thrcbr{i},\,
\end{equation}
and $i=0,1,\ldots, N_\text{max}$. The average number of repetitions for the deterministic strategy is exemplified in  Fig.~\ref{fig:retrans_strategy}. 

The number of repetitions $\Ntx$ is then deterministic and equal to $\Ntxav(\netcbr{})$.

\textbf{Probabilistic strategy.} In the \textit{probabilistic} strategy,  the average number of repetitions $\Ntxav(\netcbr{})$ follows a piece-wise linear function. Specifically, 
the station transmits an average number of repetitions equal to 

\begin{align}\label{eq:Ntx_prob}
\Ntxav(\netcbr{}) = \vInBox{\averageN(\netcbr{})}{N_\text{min}}{N_\text{max}} 
\end{align}
where $N_\text{min}=0$ and
\begin{equation}\label{eq:Navtx_prob}
\begin{split}
\averageN(\netcbr{}) &= 
k-0.5+\frac{\thrcbr{k}-\netcbr{}}{\thrcbr{k}-\thrcbr{k+1}} \text{ when } \thrcbr{i+1}\leq\netcbr{}<\thrcbr{i},\\ i&=0,1,\ldots,N_\text{max}
\end{split}
\end{equation}
where 
$k=\vInBox{i}{N_{\min}+1}{N_{\max}-1}$. 
The average number of repetitions for the probabilistic strategy is exemplified in  Fig.~\ref{fig:retrans_strategy}.

The vehicle then determines $\Ntx$ as a random variable equal to 
\begin{equation}
\Ntx = \lfloor\Ntxav(\netcbr{})\rfloor +\delta(\netcbr{}) 
\end{equation}
where $\delta(\netcbr{})$ is a Bernoulli random variable equal to 1 with probability $p(\netcbr{})=\Ntxav(\netcbr{})-\lfloor\Ntxav(\netcbr{})\rfloor$.

This definition implies that $\delta(\netcbr{})=1$ 
with probability $p(\netcbr{})=0.5$ when the \ac{CBR} value $\netcbr$ equals any of the thresholds $\thrcbr{i}$ (the orange points in Fig.~\ref{fig:retrans_strategy}). When the \ac{CBR} is between two thresholds, such probability varies linearly with $\gamma$. The piece-wise function for $\netcbr{}$ below $\thrcbr{N_{\max}}$ and above $\thrcbr{N_{\min}+1}$ maintains the slope of the adjacent intervals until reaching the maximum ($N_{\max}$) and minimum ($N_{\min}$) values, respectively. As an example, when $\netcbr=\thrcbr{2}$, the average number of repetitions is 1.5, i.e. the station sets 1 repetition with probability 0.5 and 2 repetitions otherwise. 


\section{Results and discussion}
\begin{table*}[!t]
\caption{Main simulation parameters and settings
\vspace{-2mm}
\label{Tab:Settings}}
\footnotesize
\centering
\begin{tabular}{p{2.5cm}p{14cm}}
\hline \hline
\emph{\textbf{Scenario}} & Highway, 3+3 lanes, variable vehicle density, average speed 120 km/h with 12 km/h std. deviation\\
\emph{\textbf{Data traffic}} & 350~bytes packets generated every 100 ms \\
\emph{\textbf{MAC settings}} & Maximum contention window 15, AIFS 110 $\mu s$, SIFS 32 $\mu s$\\
\emph{\textbf{MCS}} & MCS 2 (QPSK, CR$=0.5$), with error rate probability based on the curves in \cite{BazZanIoaMar:C20} (1~dB @ PER=0.5)\\
\emph{\textbf{Rx thresholds}} & For preamble detection -100~dBm, for unknown signals -65~dBm, for the CBR evaluation -85~dBm\\ 
\emph{\textbf{Channel and power}} & Single 10 MHz channel at 5.9~GHz, tx power 23 dBm (not including antenna gain), antenna gain 3~dBi, noise figure 6~dB\\
\emph{\textbf{Propagation}} & WINNER+ Scenario B1 propagation model, correlated shadowing with 3~dB variance and decorr. dist. 25~m \\
\hline \hline
\end{tabular}
\end{table*}
 
In this section, we first assess the impact of repetitions assuming that all nodes adopt the same number of repetitions, and considering different modeling of the preamble detection and propagation models. 
Then, we show the effect of both the proposed strategies assuming a fully distributed decision.
A modified version of the open source simulator WiLabV2Xsim is used \cite{TodBarCamMolBerBaz:21}.\footnote{The simulator is available at {https://github.com/V2Xgithub/WiLabV2Xsim} and the modifications will be available in future releases.}  Table~\ref{Tab:Settings} shows the main simulation settings. 


\subsection{On the impact of preamble detection}
As described in Section~\ref{sec:rep}, the packet consists of preamble and data field. The receiver tries to decode the data field only if it first detects the preamble. Otherwise, the packet is treated as noise. 
This means that the receiver is not always able to store all the copies, as assumed in related work where the preamble detection process is not taken into account. This is especially true with a small difference between the min. power to detect the preamble and the min. power to decode the packet.\footnote{Here, the sensing threshold for preamble detection is set to -100~dBm, which corresponds to -2~dB signal to noise ratio and is in good agreement with off-the-shelf devices, and the power required to have PER equal to 0.5 with no interference is -97~dBm \cite{BazZanIoaMar:C20}.} 

\begin{figure}[tp]
\centering
\includegraphics[width=0.95\columnwidth]{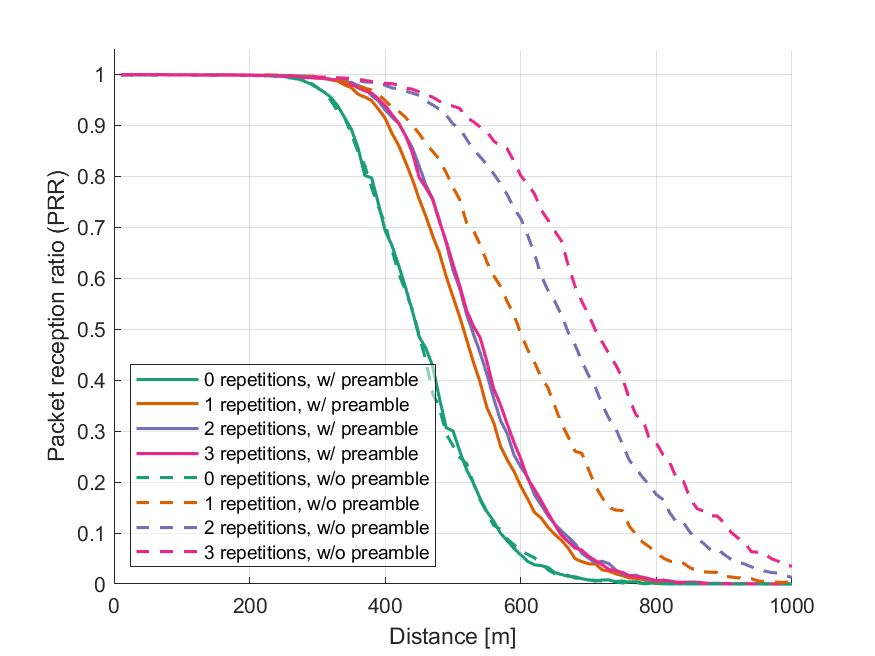}
\caption{
PRR vs. transmission distance for various number of repetitions, neglecting the preamble detection (dashed curves) or taking it into account (solid curve). Density equal to 5 veh./km.
}
\label{fig:compare_preamble}
\end{figure}

Fig.~\ref{fig:compare_preamble} shows the effect of preamble detection when the vehicle density is $5\,$veh./km. The solid and dashed lines are obtained with or without considering the preamble detection, respectively. When the distance is large (e.g., 700~m), even if more repetitions would in principle increase the overall SINR thanks to MRC (as shown by the dashed curves), the received power is mostly insufficient to detect the preamble and therefore the real improvement due to multiple transmissions is limited (solid curves). The results show that the preamble detection has a notable impact on the performance. 





\subsection{Threshold setting}
\begin{figure}[tp]
\centering
\includegraphics[width=0.95\columnwidth]{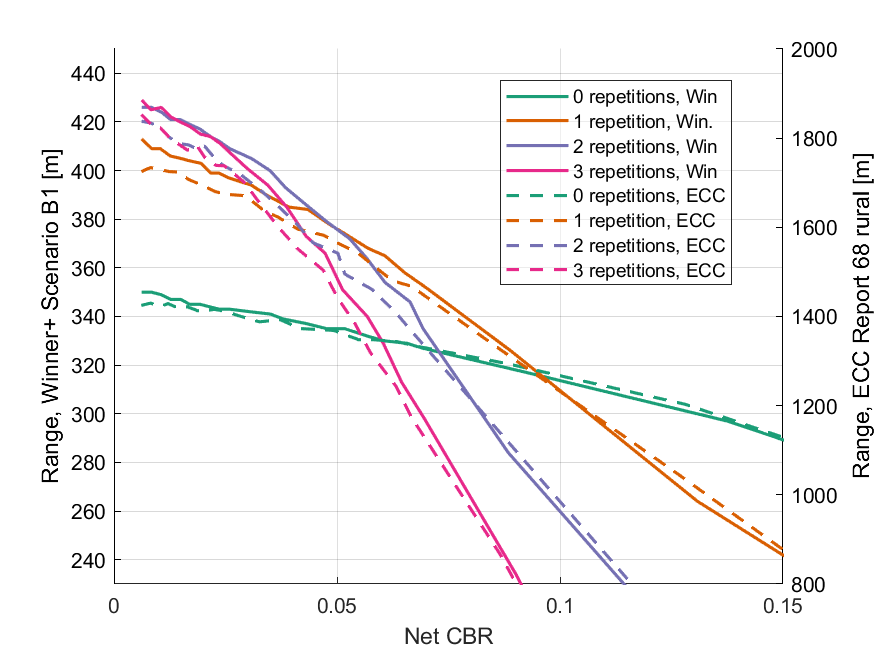}
\caption{Range 
vs. net CBR assuming different number of repetitions. Solid lines are derived from the WINNER+, scenario B1 propagation model and correspond to the left y-axis. Dashed lines are derived from the ECC rural model and correspond to the right y-axis.}
\label{fig:dis_vs_cbr}
\end{figure}

With the aim to determine the thresholds, the performance when all the vehicles adopt the same number of repetitions times chose by all vehicles at the same time is investigated in Fig.~\ref{fig:dis_vs_cbr}, where the \textit{range}, defined as the maximum distance to have $\mathrm{PRR}>0.90$, is shown varying the net \ac{CBR} $\netcbr$. 
In Fig.~\ref{fig:dis_vs_cbr}, two different propagation models are adopted to verify the generality of the derived results. In particular, in addition to the WINNER+, scenario B1 model, which is normally adopted for these kind of studies and used in the rest of the paper, also the modified ECC Report 68 rural described in \cite{etsi202110} is here considered. By comparing dashed and solid curves, results show that, apart from a scale factor on the distance, the impact of repetitions is similar for the two propagation models when looking at the net \ac{CBR}. This means that the applicability of the derived thresholds and the performance trends that follow are not limited to the specific settings adopted in this work. 

As expected, Fig.~\ref{fig:dis_vs_cbr} confirms that in a low load scenario the transmission of more repetitions improves the performance, whereas it has a negative impact when the channel load is higher. For example, if we focus on the case of net CBR larger than approximately 0.09, we note that it is preferable that all stations do not use any repetitions compared to the case where all the stations transmit one repetition (the two cases have the same net CBR by definition).

Given these results and adopting the interval setting process explained in Sec.~\ref{sec:strategies}, 
the three thresholds are set as $\thrcbr{1} = 0.09,~~\thrcbr{2} = 0.05,~~\thrcbr{3} = 0.03$.

\subsection{Performance of the proposed strategies}
\begin{figure}[tp]
\centering
\includegraphics[width=0.95\columnwidth]{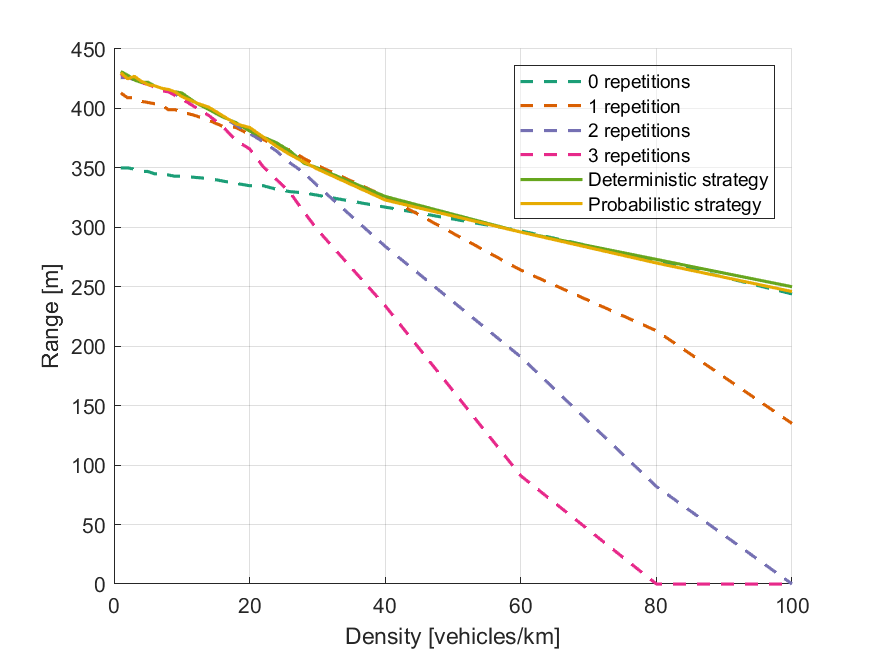}
\caption{Effectiveness of the repetition strategies. The dashed curves correspond to the cases where all the vehicles use the same and fixed number of repetitions. The solid curves correspond to the cases where all the stations autonomously adopt either repetition strategy.}
\label{fig:compare_results}
\end{figure}
Fig.~\ref{fig:compare_results} shows the range 
varying the vehicle density, assuming either all the stations adopt the same and fixed number of repetitions, or all the stations adopt one of the strategies detailed in Sec.~\ref{sec:strategies}. Please note that in the simulations the stations are completely unsynchronized and the intervals of duration $\Tcbr$ used to calculate the CBR are independent among the stations. 
It can be noted that 
both the deterministic and probabilistic strategies, for any value of the density, approximately provide the same performance as the best solution with a fixed number of repetitions, therefore demonstrating the validity of the proposed approach. 



\begin{figure}
\centering
\includegraphics[width=1\columnwidth]{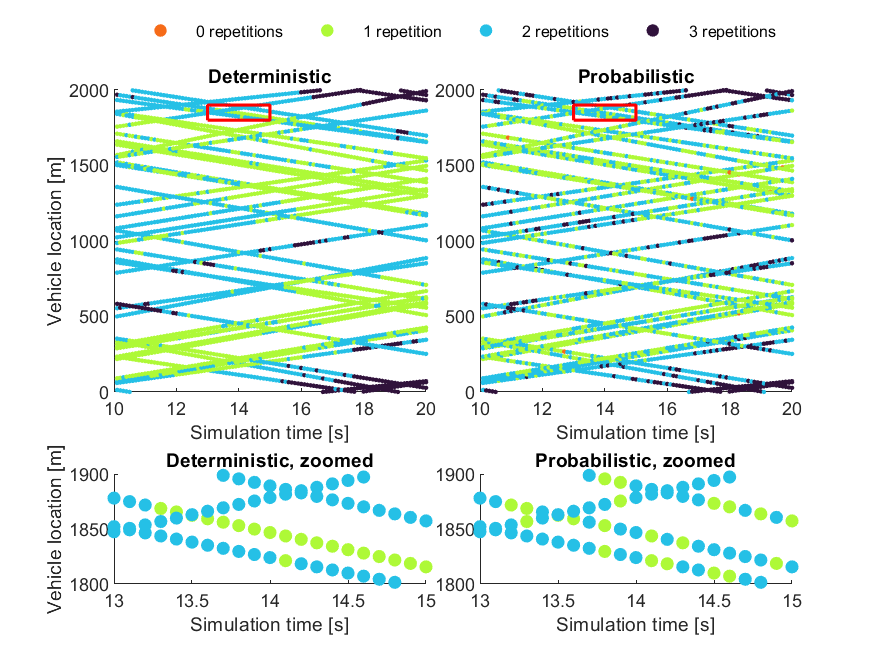}
\caption{Repetitions used by the vehicles in a sampled time interval plotted as a function of space and time. The two figures located lower side are zooms of the red boxes in the figures plotted upper side. Density equal to 20~veh./km.}
\label{fig:compare_location_allin1}
\end{figure}

Even if both strategies are effective to optimize the network performance, the probabilistic strategy allows a better distribution of the number of repetitions. This is shown through Fig.~\ref{fig:compare_location_allin1}, which illustrates the impact of the two repetition strategies over time and space in a sample simulated interval. The two bottom-side subfigures are zooms of the upper-side ones. In each subfigure, the y-axis represents the location of the stations and the x-axis the time; the colors indicate the number of repetitions set by the station located in that position at that time. The figure shows that with the deterministic strategy the vehicles tend to maintain the same number of repetitions for longer intervals than with the probabilistic strategy; for example, looking at magnified parts in the lower subfigures, with the deterministic strategy all stations keep using 2 repetitions except a single one, which keeps using only 1 and is therefore penalized; a variable and thus fairer number of repetitions is instead selected by all stations in the probabilistic case.

\section{Conclusion}
Considering the possibility added by IEEE 802.11bd to transmit more than one replica of the same packet to improve the reliability of V2X communications, in this paper we proposed two strategies to let each station set the  number of repetitions in a fully distributed way based on local measurements of the channel load, with the objective  to maximize the network performance. The effectiveness of the proposed strategies is validated through network-level simulations performed in complex scenarios where each station autonomously performs the selection under variable conditions.

\bibliographystyle{IEEEtran}
\bibliography{myBib}

\end{document}